\documentclass[useAMS,usenatbib,aps,prd,twocolumn]{revtex4-1}
\usepackage{times,graphicx,amsmath,amsfonts,amssymb,aas_macros,epstopdf,ulem}
\usepackage{epsfig,enumerate}
\usepackage[usenames,dvipsnames]{xcolor}
\usepackage{url}
\usepackage{subfigure}
\usepackage[colorlinks=true, urlcolor=black]{hyperref}
\usepackage[utf8]{inputenc}
\usepackage[]{ulem}

\def\etal{{\frenchspacing\it et al.}}
\def\ie{{\frenchspacing\it i.e.}}
\def\eg{{\frenchspacing\it e.g.}}

\def\be{\begin{equation}}
\def\ee{\end{equation}}
\def\ba{\begin{eqnarray}}
\def\ea{\end{eqnarray}} 
\def\Msun{h^{-1}{\rm M}_{\odot}}
\def\hmpc{h^{-1}\,{\rm Mpc}}

\def\kms{\, {\rm km }\, {\rm s}^{-1}}
\def\hmpcc{h^{-3}\,{\rm Mpc}^3}
\def\hkpc{h^{-1}\,{\rm kpc}}
\def\dd{\textrm{d}}
\def\lcdm{\Lambda{\rm CDM}}
\def\de{\delta}

\def\ln{{\rm ln}\,}

\newcommand{\av}[1]{\langle{#1}\rangle}

\def\frac#1#2{{\textstyle{#1\over #2}}}

\def\simlt{\stackrel{<}{{}_\sim}}
\def\simgt{\stackrel{>}{{}_\sim}}

\def\vmax{V_{\textrm{max}}}
\def\rmax{R_{\textrm{max}}}

\newcommand{\scolor}{\textsc{color}}
\newcommand{\coco}{\textsc{coco}}
\newcommand{\nexus}{\textsc{NEXUS}}

\newcommand{\subfind}{\textsc{subfind}}

\definecolor{nodes}{rgb}{1,0.27,0}
\definecolor{fila}{rgb}{1.0,0.65,0}
\definecolor{walls}{rgb}{0.0,0.25,0.0}
\definecolor{voids}{rgb}{0.0,0.0,1.0}

\begin{document}
\title{Caught in the Cosmic Web: environmental effect on halo concentrations, shape and spin.}
\author{Wojciech A.~Hellwing$^{1}$, Marius Cautun$^{2}$, Rien van de Weygaert$^{3}$ and Bernard T.~Jones$^{3}$}
\affiliation{$^{1}$Center for Theoretical Physics, Polish Academy of Sciences, Al. Lotników 32/46, 02-668 Warsaw, Poland}
\affiliation{$^{2}$Leiden Observatory, Leiden University, PO Box 9513, NL-2300 RA Leiden, the Netherlands}
\affiliation{$^{3}$Kapteyn Astronomical Institute, University of Groningen, PO Box 800, NL-9747 AD Groningen, the Netherlands}
\date{\today}

\begin{abstract}
Using a set of high-resolution simulations we study the statistical correlation of dark matter halo properties with the large-scale environment.
We consider halo populations split into four Cosmic Web (CW) elements: voids, walls, filaments, and nodes. 
For the first time we present a study of CW effects for halos covering six decades in mass: $10^{8}-10^{14}{h^{-1}{\rm M}_{\odot}}$.
We find that the fraction of halos living in various web components is a strong function of mass, with 
the majority of $M>10^{12}{h^{-1}{\rm M}_{\odot}}$ halos living in filaments and nodes. Low mass halos are more equitably distributed 
in filaments, walls, and voids. For halo density profiles and formation times we find a universal mass threshold 
of $M_{th}\sim6\times10^{10}{h^{-1}{\rm M}_{\odot}}$ below which these properties vary with environment. 
Here, filament halos have the steepest concentration-mass relation, walls are close to the overall mean, and void halos 
have the flattest relation. This amounts to $c_{200}$ for filament and void halos that are respectively 14\% higher and 7\% lower 
than the mean at $M=2\times10^8{h^{-1}{\rm M}_{\odot}}$.
We find double power-law fits that very well describe $c(M)$ for the four environments in the whole probed mass range. 
A complementary picture is found for the average formation times, with the mass-formation time relations following trends 
shown for the concentrations: the nodes halos being the oldest and void halo the youngest. The CW environmental effect 
is much weaker when studying the halo spin and shapes. The trend with halo mass is reversed: the small halos 
with $M<10^{10}{h^{-1}{\rm M}_{\odot}}$ seem to be unaffected by the CW environment. Some weak trends are visible 
for more massive void and walls halos, which, on average, are characterized by lower spin and higher triaxiality parameters.    
\end{abstract}

\maketitle
\section{Introduction}
\label{sec:intro}

The standard model of cosmology is very successful in explaining an impressive number of observations, spanning a vast 
range in time and space from primordial nucelosynthesis to the low-redshift 
large-scale distribution of galaxies. 
The latter represents the magnificent Cosmic Web: a large-scale network of 
galaxy clusters as nodes from which cosmic filaments spread out, which in turn act as the scaffolding 
for cosmic sheets that bound vast, nearly empty, voids. The success of the standard cosmological model in explaining 
observations regarding early Universe and large-scale structure statistics is undisputed, but the problem of explaining, 
in full details, the observed population of galaxies and the dark matter halos they live in, is still an 
open one. Galaxy formation is a separate theory, which takes the background cosmological model only as an input 
and as such can be regarded as an independent problem. The large-scale structure and its most prominent manifestation 
in the form of the Cosmic Web serves as the natural environment in which DM halos and galaxies are 
formed and nurtured \citep{Bond1996,Weygaert2009,Cautun2014}.
The connection and interplay between the large-scale structure environment and 
some intrinsic properties of DM halos and galaxies is a subject of growing interest and study 
\citep[\eg][]{AragonCalvo2007,Hahn2007,vdWeyagert2008,AragonCalvo2010,Bond2010,Alpaslan2014,Alonso2015,Leclercq2015,Metuki2016,Pomarede2017,Libeskind2018,GV2018,Martizzi2019,Xu2020}.

In the past two decades, the various studies of both simulation and the observational data revealed
that the Cosmic Web, or more broadly the Large-Scale Structure environment in which halos and galaxies
are embedded, affects a number of properties of these
objects. It was found that the local vorticity and tidal fields play important role in galaxy and halo
spin acquisition (both magnitude and direction)
\citep[\eg][]{Heavens1988,Robertson2006,Maccio2007,Jones2010,Libeskind2013,Wang2018,GV2019,GV2020,Chen2020}.
The Cosmic Web environment affects halos and galaxies in many aspects, from shaping the anisotropic distribution
of satellite galaxies \citep{Libeskind2015,shishao2016,Shao2018}, to halo concentrations and their assembly
histories \citep{Avila-Reese2005,Rey2019,Chen2020,WangKuan2020}, and halo shapes \citep{GV2018,Despali2017,Lau2020}.
The large-scale structure environment correlates also with galaxy stellar mass\citep{Kauffmann2004,Metuki2015,Beygu2016} 
and morphological properties \citep{Wu2009,Das2015,Darvish2017,Poudel2017,Wei2017,Sazonova2020,Miraghaei2020}.

A clear example on how DM halos are affected by their large-scale environment 
is the so-called assembly bias\citep{Li2008,Dalal2008,Zentner2014,Borzyszkowski2017,Busch2017,Contreras2019}. Here,
the older halos are found to be more clustered than the universal mean which highlights that the older halos
are preferentially found in denser parts of the large-scale density field \citep{Croton2007,Gao2007}.

From the observational point of view a plethora of measurements clearly indicate that the Local Group and its
most immediate cosmic surrounding stretching up to a hundred megaparsecs creates a unique Local Universe
ecosystem \citep{Elyiv2013,Norma,Vela}. Here, the Local Void \citep{Karachentsev2012}, Virgo and Coma clusters 
act as major agents that shape the dynamics of nearby galaxies \citep{Hellwing2017,Hellwing2018} 
and drive the build-up and assembly of the local galaxy population
\citep{Garrison-Kimmel2014,Brook2014,Read2017,Desmond2018,Carlesi2020}. 
Such environmental effects are also seen in even larger volumes such as those of deep galaxy redshift surveys 
\citep[\eg][]{Eardley2015,Brouwer2016,Pandey2020}.

The DM halos are primary hosts for galaxy formation \citep{WhiteRees1978}, 
where baryons condense to form stars that eventually grow into galaxies. The fundamental properties
of DM halos, such as their total virial mass or internal density distribution set the time and length
scales for the galaxy formation physics. Thus these elementary properties and their time evolution
impact the galaxies that they host \citep{SAMSCole2000,MS,Springel2006}. Our modelling and understanding 
of galaxy formation is rooted in the original model of Ref.~\citep{WhiteRees1978}, which has been
extensively studies and significantly extended since its initial formulation 
\citep[\eg see][]{DekelSilk1986,White1991,Kauffmann1993,Cole1994,Mo1998,SAMSCole2000}. 
In this model, galaxies form in the centre of a halo,
and when accreted into a bigger halo become satellites. By its nature, 
this framework relies mostly on local dark matter and gas properties with minor contributions
from the scales larger than the halo itself. In this model DM halos are fully self-similar.
To move beyond this simplified picture and account for environmental effects a careful assessment of the
the Cosmic Web impact is needed.

The Cosmic Web is usually characterized into four distinct morphological types or environments: voids, walls,
filaments and clusters; there is no doubt that specific localisation in such network determine which and to what
extend halo properties will be affected. Velocity and density profiles of DM halos are a primary input 
for analysis and interpretation of various observations: orbital kinematics, masses and gravitational potential, the
gamma-ray annihilation signal, strong and weak lensing observations, etc. \citep[for a review see][]{FrenkWhite2012}. 
Thereby, identifying which halo properties are significantly correlated with the Cosmic Web environment and assessing
some average trends for their populations will enable a proper inclusion of such environmental effects into 
modelling of both halo and galaxy-based observables. This constitute the main goal of this work. 
In this paper we study the dependence of a multitude of halo properties, such as density 
profiles, assembly times, shape and spin, on the large-scale Cosmic Web environment.
The emphasis is put on halo mass-concentration relation, which is an important ingredient in many theoretical models
and its connection to the average halo formation times. We assess how the Cosmic Web, on average, affects those properties
in a systematic way.

This paper is organized in the following way: in \S\ref{sec:simulation} we describe our input data sets from
a number of numerical simulations; in \S\ref{sec:nexus_CW} we present the Cosmic Web identification algorithm
of our choosing. The section \S\ref{sec:results} we present the main results of our analysis, which is followed
by concluding remarks given in \S\ref{sec:conclusions}.

\section{Simulations}
\label{sec:simulation}

In this work we use a suite of very high resolution N-body simulations: {\it COpernicus COmplexio} (\coco)
and {\it COpernicus complexio LOw Resolution} (\scolor)\citep[see more in][]{COCO1,COCO2,COCO3}.
The first is a zoom-in simulation representing 
at $z=0$ a roughly spherical region encompassing a volume of $V_{hr}\approx 2.\times 10^{4}\hmpcc$
(a sphere with an effective radius $R_{hr}\approx 17.4\hmpc$). The high-resolution \coco~ region is embedded in
a larger uniform lower resolution box, $70.4\hmpc$ on a side. \scolor~ is the parent simulation, from which the zoom-in 
region is drawn from. Thus, effectively the \coco~ is a fine-sampled sub-volume of the \scolor~ box. The \coco~ simulation
consists of 12.9 billion of high-resolution particles ($\sim 2340^3$), each with $m_p=1.135\times 10^5\Msun$.
The parent \scolor~ is a set-up of $1620^3$ particles sampled with mass resolution of $m_p=6.2\times10^6\Msun$.
The cosmological parameters used to set the initial power spectrum of matter fluctuations and to fix the expansion
history were those of the seventh year result from {\it Wilkinson Microwave Anisotropy Porbe} (WMAP) \citep{WMAP7}:
$\Omega_{m0}=0.272$, $\Omega_{\Lambda0}=0.728$, $\Omega_b=0.04455$, $\Omega_k=0$, $h=0.704$, $\sigma_8=0.81$
and $n_s=0.967$.

We use the original halo and subhalo catalogs from \coco~ and \scolor~ samples identified by the \subfind{} algorithm \citep{SUBFIND}.
\subfind{} begins by identifying DM groups using the friends-of-friends (FOF) algorithm \citep{Davis1985}, a standard linking length 
of $b=0.2$ times 
the mean inter particle separation was used. All FOF groups  with at least 20 particles were kept for further analysis. 
Next the algorithms analyses each FOF group to find gravitationally bound DM subhalos (i.e. substructures within the FOF halos).
Potential subhalos are first marked by searching for overdense regions inside the FOF groups that are next pruned by keeping only those
particles that are gravitationally bounded. This results in a catalog of self-bounded structures containing at least 20 particles.
For each halo and subhalo we also compute and store a number of additional properties. 

The FOF groups (or halos as we will call them interchangeably) are characterized in terms of their FOF mass, $M_{FOF}$, 
as well as of their $M_{200}$ mass. The first is given by the mass contained in all the particles associated to a given FOF group.
In contrast, $M_{200}$ is the mass contained in a sphere of radius $r_{200}$ centered on the FOF group, such that
the average overdensity inside the sphere is $200$ times the critical closure density, $\rho_c\equiv (3H(z)^2)/(8\pi G)$.
With $G$ indicating Newton's gravitational constant and $H(z)$ being the Hubble parameter.

The halo and subhalo catalogs obtained using the above described procedure 
are further analyzed and post-processed to compute a number of internal properties, such as 
density profiles and shape and spin parameters. We discuss the specific details in the relevant sections below.

The halo merger trees are constructed for both simulations run using an updated algorithm 
that has been developed for use with the semi-analytic galaxy formation code \verb#GALFORM# \citep{SAMSCole2000}. The method we 
used is described 
in detail in \cite{Mtrees3} and more details of the merger trees of the \coco{} and \scolor{} simulations can be found in
the original simulation paper \cite{COCO1}. The essential part of the algorithm consists of unique linking between
subhalos from two consecutive snapshots.
In this analysis we will be mainly interested in following a halo's most massive progenitor in the tree, which is simply obtained by
walking the tree along the most-massive branch.

\section{NEXUS Cosmic Web}
\label{sec:nexus_CW}
\begin{figure*} 
    \includegraphics[angle=0,width=0.48\textwidth]{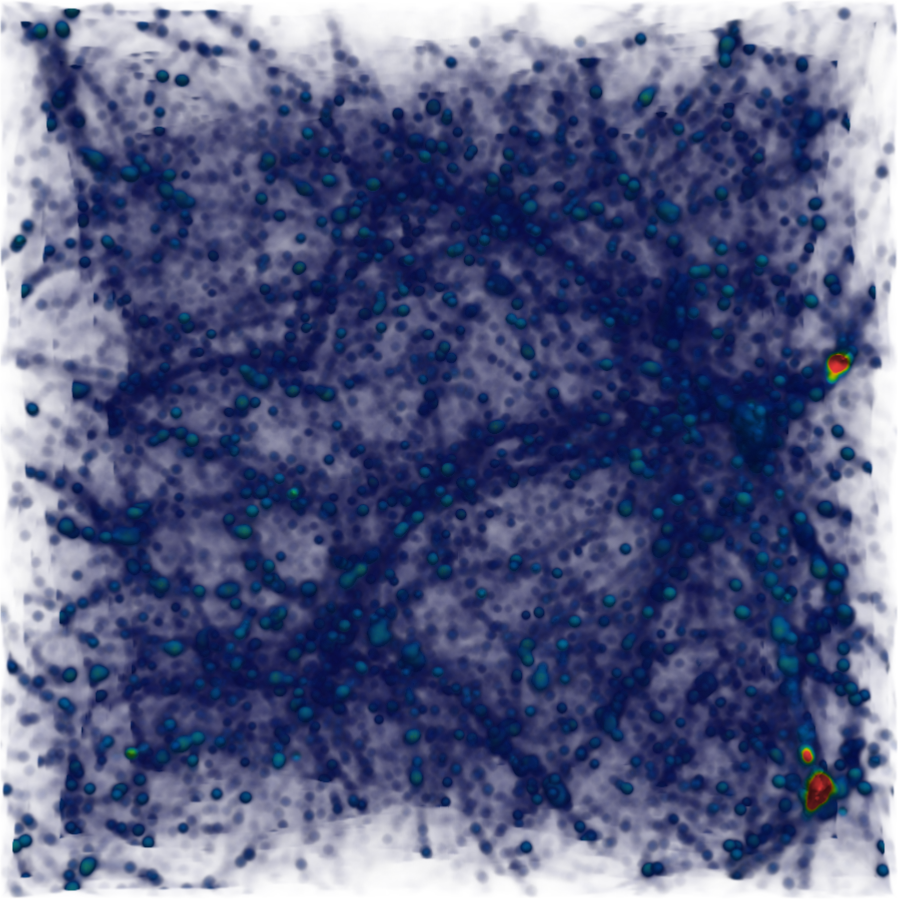}
    \includegraphics[angle=0,width=0.48\textwidth]{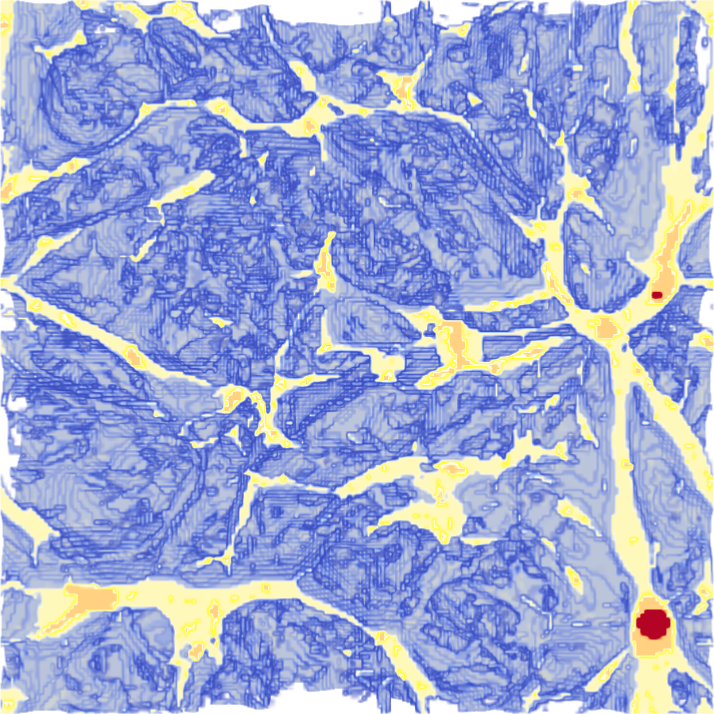}
    \caption{ The DTFE density rendering (on the left) and the corresponding \nexus+ Cosmic Web map (on the right).
    The two panels show the \coco/\scolor{} box with $70.4\times70.4 \hmpc$ on a side with a perspective z-direction mapping.
    The density is rendered so the below average density regions (all cosmic voids and some walls) are transparent. 
    The corresponding Cosmic Web map on the right is rendered by setting all void regions to be 
    fully transparent and the cosmic walls (i.e. void boundaries) 
    are displayed in a tomographic projection with light shadows on a blue surface. 
    On the near face of the simulation volume we see cross-sections through the nodes in red,
    filaments in orange and the walls in yellow. The nodes and filaments cannot be seen at
    farther distances since they are surrounded by thin wall regions. 
    }
    \label{fig:CWdensity}
\end{figure*}

We use the \nexus+ algorithm \citep{Cautun2013} for the segmentation of the Cosmic Web into its distinct morphological components: 
{\bf \textcolor{nodes}{nodes}}, {\bf \textcolor{fila}{filaments}}, {\bf \textcolor{walls}{walls}}
and {\bf \textcolor{voids}{voids}}. This method is an improved version of the MultiScale Morphology Filter \citep{AragonCalvo2007} and includes 
more physically motivated prescriptions for determining the web environments. 
We chose this method due to its multiscale and parameter free character that make it an ideal tool for identifying
in a robust manner the cosmic environments. Due to its scale-space approach, this method is equally 
sensitive in the detection of both prominent and tenuous filaments and walls. The tenuous environments are especially important 
to obtains a complete 
census of web environments for the faintest galaxies (i.e. lowest mass halos), since many of them are found in tendrils 
criss-crossing the underdense 
regions \citep{Cautun2014,Alpaslan2014}.

The \nexus+ algorithm takes as input the DM density field on a regular grid. To take full advantage of the multitude of structures resolved by 
the COLOR simulation, we use the Delaunay Tessellation Field Estimator (DTFE) \citep{Schaap2000,Weygaert2009} to interpolate the density to a $640^3$ 
grid corresponding to a $0.11\hmpc$ grid spacing. \nexus+ starts by smoothing the input density field on a suite of scales from $0.125$ to $2\hmpc$. 
For this, it uses the Log-density filter, which corresponds to a Gaussian smoothing of the logarithm of the density field \citep{Cautun2013}. 
For each smoothing scale, 
the resulting density is used to calculate the Hessian matrix and its three eigenvalues. The values and signs of the eigenvalues are used to determine 
the environment response at each location, i.e. grid cell. The actual expression is rather involved (see \cite{Cautun2013}), but qualitatively a region 
is classified as a filament if the two largest eigenvalues are negative (i.e. indicate collapse around those directions) and if their absolute values are much larger than 
the third eigenvalue. Then, at each location the results of all smoothing scales are combined by taking the maximum of all the values. This is 
because a web structure of a given thickness has its largest web signature when smoothing with a filter of the same width. Finally, the nodes, 
filaments, and walls are identified as all the grid cells whose environment response is above a self-determined threshold value 
(see \cite{Cautun2013} for details). 
The remaining volume elements that are not associated to nodes, filaments, or walls, are classified as part of voids.
Our final Cosmic Web map consists of $256^3$ cells (i.e. $275\hkpc$ grid spacing) where each cell is assigned one of the four web environments.

The \nexus+{} web environments are uniquely and robustly defined for an input density filed.
The algorithm assigns to a given region of space the flag of an environment that has the strongest response function over a set of smoothing scales.
It has been shown \citep{Cautun2013} that the resulting environment classifications goes
beyond a simple local-density mapping, and is sensitive to a multi-scale hierarchical nature of the Cosmic-Web. 
Because of this feature, the \nexus+ environments only partially can be characterised by their density distribution functions,
and the resulting PDFs are largerly overlapping \citep[see Fig. 4 in][]{Libeskind2018}.
\begin{figure*} 
    \includegraphics[angle=0,width=0.48\textwidth]{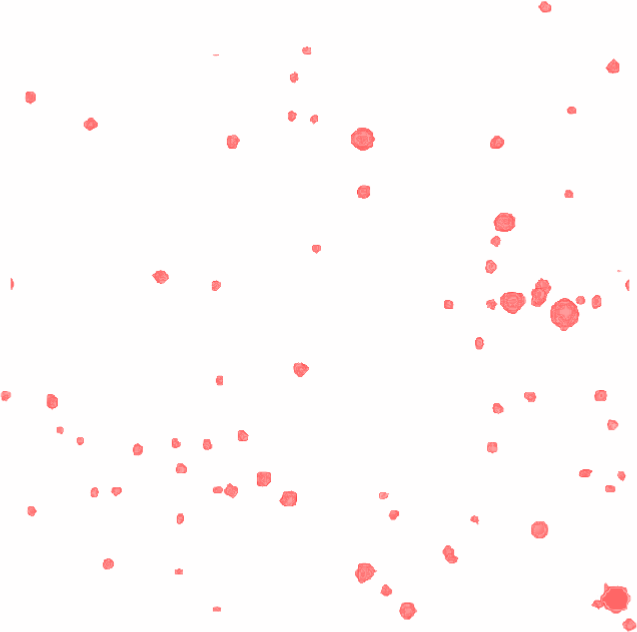}
    \includegraphics[angle=0,width=0.48\textwidth]{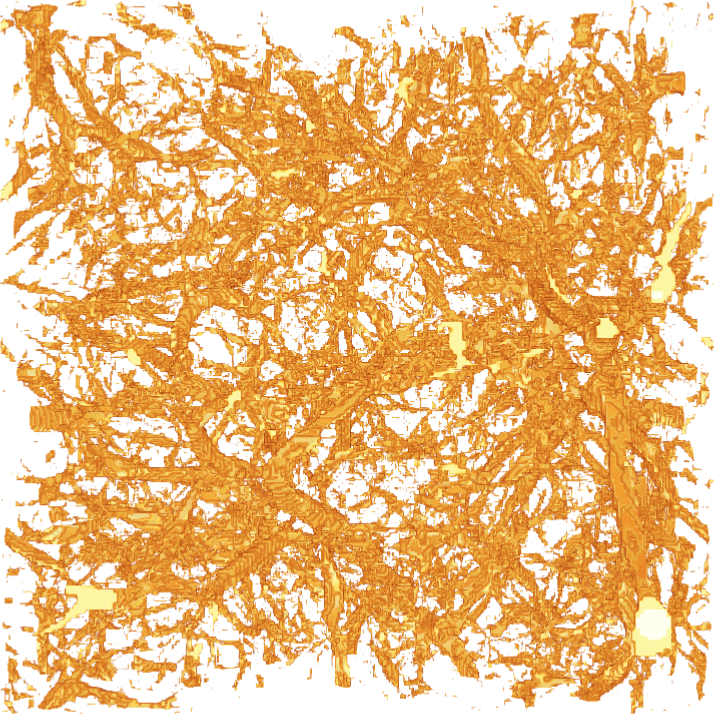}\\
    \includegraphics[angle=0,width=0.48\textwidth]{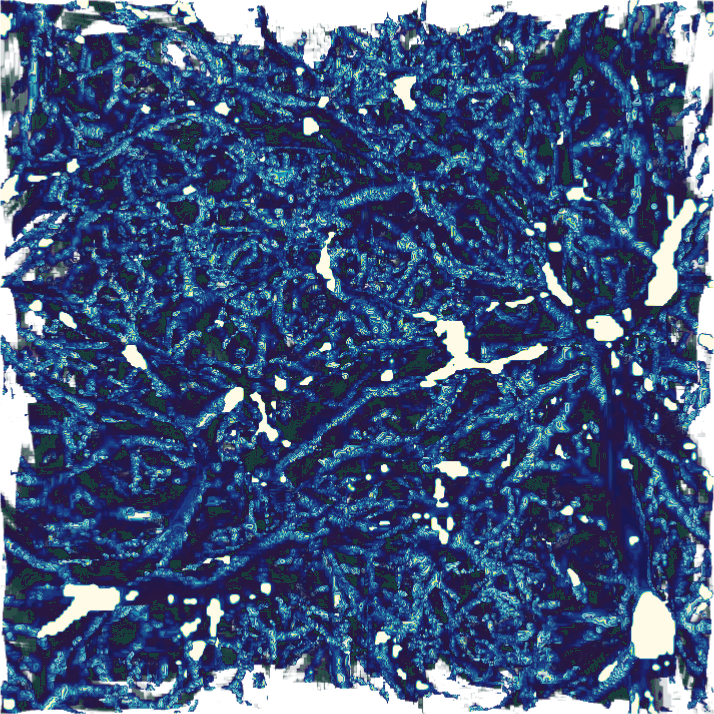}
    \caption{ The \coco/\scolor{} volume \nexus+{} maps showing nodes (the upper-left panel), filaments (the upper-right)
    and walls (the bottom panel). The magnificent level of detail with which we recover the Cosmic Web is a result of the
    very high resolution of the \coco/\scolor{} suite.
    }
    \label{fig:Coco_nexus}
\end{figure*}

\section{Results}
\label{sec:results}
In what follows we present and discuss our main results. We use three different estimators to measure
our sample statistical uncertainties. Where simple number counts dominate the statistics like in cases such as the mass functions 
estimations, we use the standard Poisson error. Whenever we estimate trends or functions of the data, 
we use bootstrapping to measure the uncertainty associated to the mean and median values for each bin. 
Finally, sometimes we study the spread or width of a distribution, which we quantify using $16'$th and $86'$th percentiles. 
We use percentiles instead of the usual standard dispersion since some of our data has non-Gaussian 
distributions and the standard deviation in that case could be misleading.

In FIG.~\ref{fig:CWdensity} the DTFE density together with the corresponding \nexus+ environmental map is shown. While the DTFE 
density field is continuous and is displayed using a projection rendering, the Cosmic Web map is a discrete tessellation of the simulation 
space and is represented in a tomographic projection. For the density field projection we set all regions with density less then the cosmic mean 
(\ie{} $\delta<1$) to be transparent, which reveals the prominent network of filaments and nodes. We can also distinguish a number of 
spherical, ball-like regions, spread-out across the filamentary network. These indicate the prospective location of massive dark matter halos. The three 
green-red regions close to the right-hand density map boundary are nodes with surrounding thick filaments close to the projection plane.

\begin{figure}
\includegraphics[angle=0,width=0.48\textwidth]{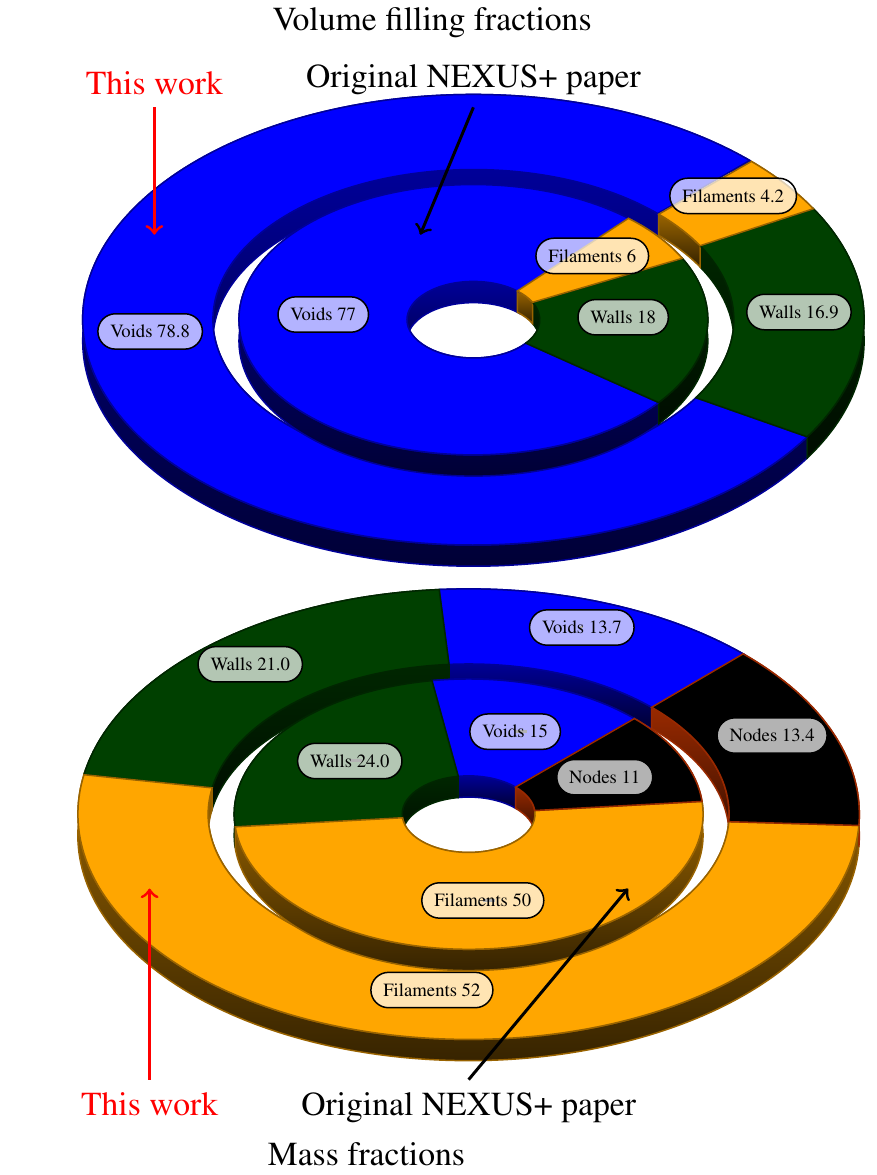}
    \caption{The Cosmic Web components in the \coco/\scolor{} simulation. 
    {\it The upper panel:} The volume filling fraction of voids (blue), walls (Green) and filament (yellow). The volume fraction 
    of the node environment, $0.045\%$,
    is so tiny that is is not displayed. {\it The bottom panel:} The fraction of the total mass found in each Cosmic Web environment. 
    Here the nodes, denoted
    by orange, constitute a significant fraction of the mass.
    }
    \label{fig:env_rat}
\end{figure}

The corresponding Cosmic Web segmentation shown in the right panel of FIG.~\ref{fig:CWdensity}
reveals a complicated inter-connected network full of small-scale details on the cosmic walls surfaces
encompassing voids. The tomographic projection is not optimal to show the filamentary network because most 
filaments are surrounded by a thin wall-like environment. To allow for a clearer visualization of the various web environments, we
plot the nodes, filaments and walls in three separate panels in FIG.~\ref{fig:Coco_nexus}.

We summarize the detailed segmentation studied here by illustrating the volume and mass-filling fractions of each Cosmic Web environment identified in \scolor{}. 
This is depicted in Fig.~\ref{fig:env_rat}.
We can compare our results (outer ring) with the reference \nexus{} results of Ref. \cite{Cautun2014}, which were obtained using the lower resolution Millennium 
simulations \cite{MS,MS2}. We find good agreement, which is indicative of the fact that the \scolor{} volume, while being smaller than the Millennium simulation one, 
it is still representative of the large-scale distribution of matter. The main difference is that while our filaments occupy slightly less volume they contain 
a somewhat higher mass fraction than the Ref. \cite{Cautun2014} filaments. This effect is due to the higher resolution of the \scolor{} simulation, which allows for
a better identification of the edges of thick filaments, hence the lower volume, and for the recovery of tenuous filaments that criss-cross the underdense regions, 
which lead to a somewhat higher mass fraction.  

\subsection{Halo mass functions}
\label{subsec:HMF}
\begin{figure}
    \includegraphics[angle=-90,width=0.48\textwidth]{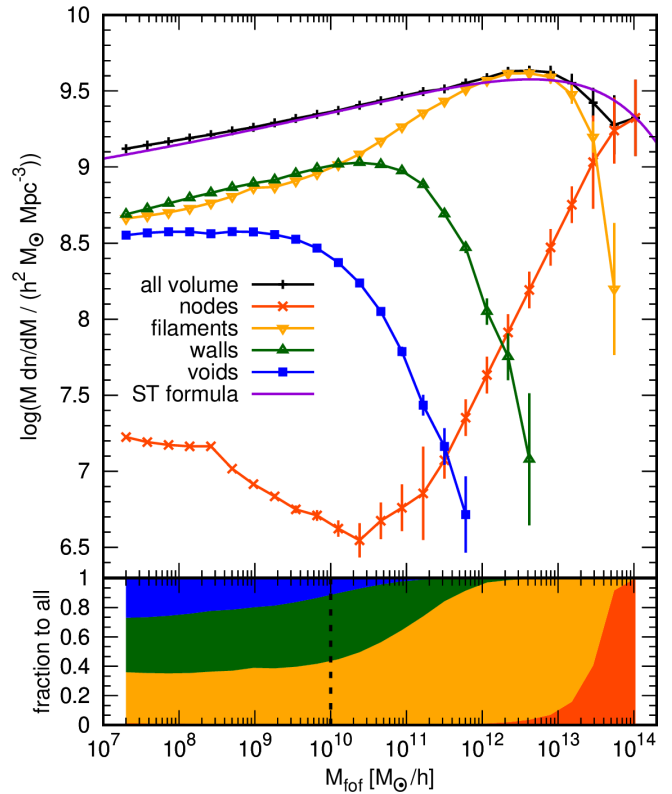}
    \caption{\coco+\scolor~ halo mass functions for different environments. {\it Upper panel:} The FOF mass functions for $z=0$. 
    The black line with crosses is the result for the whole volume, orange line with x-es depicts halos found in nodes, yellow 
    with down-triangles is for filaments, green with up-triangles marks halos in cosmic walls and finally blue line with boxes 
    shows void halo mass function. The solid magenta line illustrates the Sheth-Tormen prediction. 
    The vertical bars indicate Poisson error-bars. {\it Bottom panel:} The fractions of FOF halos found in each specific 
    environment. Lines show ratios of FOF mass functions in each environment w.r.t. the whole unsegmented simulation volume. 
    The dashed vertical line marks the mass resolution limit of the original \nexus+{} papers \citep{Cautun2013,Cautun2014}.}
    \label{fig:HMF}
\end{figure}
A fundamental characteristic of any population of halos is their mass function, often refereed to as the halo mass Function (HMF).
The HMF is simply a comoving number density of halos expressed as a function of their mass. The Cold Dark Matter (CDM) models
predict that HMF has an exponential cut-off at large masses (\ie{} cluster scales) and is characterized by a power-law, 
$\dd n/\dd M\sim M^{-\alpha}$, at small halo masses, with a slope $\alpha\sim 1$. This is due to a nearly scale-invariant
power spectrum of the primordial density fluctuations and hierarchical nature of the structure formation process of CDM models \citep{Press-Schechter,NFW2}.

The results of excursion-set modeling based on spherical collapse of the formation of halos, 
which builds upon the Press-Schechter formalism\citep{Press-Schechter}, suggests that the HMF
has an universal character when expressed in the terms of the root mean square variance of matter fluctuations, $\sigma(M,z)$, 
at given mass scale $M$ and redshift $z$.
The predictions of Press-Schechter formalism further improved by ellipsoidal-collapse models \citep{ShethTormen2002,Corasaniti2011b} (\ie{} models relaxing the assumed 
sphericity of all halos) and  has been a subject of intensive study and scrutiny in the past few decades \citep{Tinker2008,Desjacques2008,Maggiore2010,Corasaniti2011}. 
The literature about the halo mass function is very rich,
which reflects its pivotal role in cosmology. The comoving number density of dark matter halos,
as described by the HMF, is not directly observable, but it is a crucial ingredient in modeling many statistics and observables of crucial
importance in cosmology. Precise HMF is needed in semi-analytical galaxy formation models \citep{SAMSCole2000,Mtrees1}, in the halo-model of the non-linear 
matter power spectrum, galaxy clustering predictions, weak-lensing shear and convergence observations, high-redshift Lyman-$\alpha$ 1D power
spectrum and many others \citep[\eg~ see][]{Cooray2002,Conroy2006,SKibba2009,Reid2012}.

The HMF was also studied in relation to the Cosmic Web, with most results pointing toward a picture, where fractions of halos
forming and residing in different LSS environments varies as a function of their mass \citep[see \eg][]{Alonso2015}. Thus, one can decompose the overall 
HMF into different components representing the contribution of specific environments. Hence, as a starting point of our analysis,
we study the halo mass function and its decomposition into the four components of the LSS as identified by the \nexus+{} method. The \coco{} run
has a relatively small volume and by selection does not contain any cluster-mass halos. This result in a relative scarcity of massive halos
(\ie{} $M\geq10^{12}\Msun$) when compared to the expected universal mean. For that reason, to avoid any limited volume and density-related
biases we choose to use the HMFs from the regular \scolor{} box in the mass regime $3\times 10^{8}\Msun\leq M_{FOF}\leq 10^{14}\Msun$ 
and supplement it by the \coco{} sample only at the low-mass $\simlt 3\times 10^{8}\Msun$, where \scolor{} results are affected by resolution.

In FIG.~\ref{fig:HMF}
we show two panels that illustrate how the HMF in the \coco+\scolor{} sample depends on the Cosmic Web environment.
The {\it upper-panel} shows the comoving HMF
for all simulation volume, its segmentation into specific environments and, 
the Sheth-Tormen prediction
\citep{ShethTormen2002} obtained for the relevant set of cosmological parameters. Here, we use Friends-of-Friends as a halo mass, 
$M_{FOF}$, to allow a comparison with the previously published \coco{} results. The {\it lower-panel} illustrates the fraction of halos found in
each of the LSS environments in a given mass bin. The environmental HMF presented in this way enjoys 
a few interesting features. First, all of them, except for the nodes, exhibit exponential cut-offs at the high-mass end. 
These cut-offs are somewhat similar to the same feature of the all-volume halo mass, but they appear (i) sharper and 
(ii) manifest themselves at lower mass, and this mass is a specific function of the environment. 

The void HMF
has the smoothest cut-off and it is starting to plummet at $M_{FOF}\sim10^{10}\Msun$,
the abundance of wall halos is declining a bit sharper, 
and this takes place at nearly two decades higher mass of $\sim 10^{12}\Msun$ and finally the filament HMF is the one with the sharpest 
cut-off appearing at $M\sim\times10^{13}\Msun$. 
The qualitative behaviour of our environmentally-segregated HMF agrees for $M\geq10^{10}\Msun$ with
the original \nexus+{} paper results. However, here thanks to our combined \coco+\scolor{} samples we are able to study for the first time
the HMF of web environments below this mass. The fractions of small-mass halos residing in each of the tree main environments follow the trends
hinted at higher masses, with the filament ratio gradually falling and the void ratio gaining at its expense. However, two observations
are quite outstanding in this picture. First, we note that starting from masses $\simlt 5\times10^{9}\Msun$ the wall and filament 
ratios roughly equalise and then they gradually decrease together in the favor of the void environment. Secondly,
at our low mass-end, $M{\sim}2\times 10^{7}\Msun$, the division of halos among the three environments approaches an equal 
ratio of $1/3$rd
for each. This will have significant repercussion, as we will see later, since the internal properties of halos in each of 
the environments do differ noticeably at those small masses. 
We need to be cautious of numerical effects close to the minimum \coco{} halo mass for our study, which here is 
$M_{min}=100\times m_p^{COCO}=1.145\times 10^{7}\Msun$. The study of \cite{Ludlow_conv2019}, as well as the original \coco{} simulation paper 
indicate that at this limit, the $M_{FOF}$ HMF is converged to within $5\%$, while $M_{200}$ based HMF converges to $<10\%$.
The fractional contributions seen in FIG. \ref{fig:HMF} are however much larger than the uncertainty due to limited resolution, and thus unlikely 
to be significantly impacted by numerical effects.

The fact that our results suggests that at sufficiently low halo masses the fraction of halos found in voids, walls, and filaments becomes
comparable and tentatively suggest an even partition of 1/3 in each of the three environments, reflect the superior resolution of the \coco{} simulation.
This allowed for the identification of the Cosmic Web environments in voxels of side-length of only $275\hkpc$.
At this level of details, the rich internal substructure of the Cosmic Web is revealed, and we can identify the filamentary tendrils and the tenuous walls 
that criss-cross voids and divide them into sub-voids. 

An important question in the field is to what extent the variation of the HMF with the web morphology is driven 
by the change in density between environments. It seems that when defining the environment on large, $\sim10\hmpc$, 
scales the HMF is mostly determined by the density \citep[\eg][]{Alonso2015}.
However, the likely environment relevant for halo growth is the one on scales similar to the Lagrangian patch from which 
an object formed. When studying these scales, the anisotropies of the tidal field seemed to be more important 
with studies \citep[][]{Borzyszkowski2014,Paranjape2018,Ramakrishnan2019} 
showing that the anisotropy, and not the density, is the main driver of halo assembly bias. 
The multiscale nature of NEXUS+ allows it to adaptively determine the local scale on which the web is most 
pronounced and thus to capture the tidal field anisotropy on the relevant scale.
The fact that vastly different mass fractions and volume fractions of
each environments conspire to give comparable fraction of small-mass halos living in each of them is both interesting and surprising.
These merit further and deeper investigation which, however, is beyond the scope of the current paper and we postpone it for future work.

\subsection{Density profiles}
\label{subsec:density_profiles}

We have shown already that at sufficiently small masses (\ie{} $\simlt10^{10}\Msun$) the dark matter halo population segments into three
environmental components, with significant fractions of halos locked in walls ($\simlt 35\%$), voids ($\simgt 25\%$) and 
filaments ($\simlt 35\%$).
Now, we will investigate how the various web components affect the internal properties of their inhabiting halos. 
More precisely, we will compare averages over halo populations divided into each environments 
binned as a function of their $M_{200}$ mass.

Our main focus now is on the halo density profiles. Understanding their average trends with mass and evolution with time is of major
importance for modern cosmology. We shall describe the halo density profile using the universal Navarro-Frenk-White 
(NFW) profile \citep{NFW1, NFW2}, which has been
shown to be a reasonably good description of spherically averaged halo mass distribution for the majority of DM halos 
\citep[but see also \eg][]{Neto2007,Ludlow2016,Wang2020}. 
We use the NFW profile parameterized as 
\be
\label{eqn:NFW-p}
{\rho(r)\over\rho_c} = {\delta_c\over{r/r_s(1+r/r_s)^2}}\,,
\ee
where $\rho(r)$ is the halo density averaged in a spherical shell of radius $r$, $\rho_c$ is the critical density of a flat universe,
$\delta_c$ is the halo inner characteristic overdensity and $r_s$ is the so-called scaling radius.
A standard approach that simplifies the analysis of the halo profiles is to define the concentration, $c_{200}$, parameter, expressed as:
\be
\label{eqn:concentration}
c_{200} = {r_{200}\over r_s}\,.
\ee
With the concentration parameter defined,
the NFW profile effectively becomes a one parameter fit. This is becasue, the characteristic overdensity can be now expressed as:
\be
\label{eqn:delta_c-c}
\delta_c={200\over 3}{c_{200}^3\over \ln(1+c_{200})-c_{200}/(1+c_{200})}\,.
\ee

The NFW scale radius, $r_s$, gives the radial position at which the $r^2\rho(r)$ curve attains its maximum, which sometimes is also 
denoted by $r_{-2}\equiv r_s$. For the majority of DM halos, the peak of the $r^2\rho(r)$ curve is relatively broad. 
This means that, for halos resolved with a relatively small number of particles, the exact location $r_{-2}$ of the peak 
is subject to significant
uncertainties due to the presence of noise. This, however can be significantly reduced when one works with stacked density profiles.

In our analysis we are interested to study $c_{200}$ for our halo populations split across different Cosmic Web environments. 
To get this we fit 
the NFW profile to all the halos with at least $N^{min}_p=1200$ particles for both \coco{} and \scolor{} samples separately. 
This yields  $M^{min}_{200}=1.6\times10^{8}$ and $7.4\times10^{9}\Msun$ for \coco{} and \scolor{} simulations, respectively. 
For the specific discussion of the fitting procedure we refer the reader to the original simulation paper \citep{COCO1}.

Finally, as already mentioned the NFW profile is a good fit for objects that are close to dynamical equilibrium.
Halos affected by recent major mergers or close encounters usually are not in equilibrium and the NFW functional form is not a good fit to their
mass profiles. As a consequence, the concentration  parameter derived from  fitting non-relaxed halos is ill defined
and at best biased low \citep[see \eg][]{Neto2007, CMGao2008, Ludlow2010}. To overcome this  problem, we remove 
non-virialized halos,
i.e. objects that do not jointly satisfy the following three criteria \citep{Neto2007}:
\begin{enumerate}[i]
    \item the fraction of halo mass contained in its resolved substructure is $f_{sub}<0.1$,
    \item the displacement between the center of mass and the minimum of the gravitational potential cannot exceed $7\%$ of halo's virial radius, $r_{200}$,
    \item we require that the adjusted virial ratio $K_{vir}\equiv(2T-E_s)/|U|$, is $K_{vir}<1.35$.
\end{enumerate}
Here $T$ and $U$ are the halo's total kinetic and potential energy and  we include the Chandrasekhar's pressure  term, $E_s$, 
which quantifies 
the degree to which a given halo interacts with its surroundings. See \cite{Shaw2005} and their eqn.~(6) for the definition and method used 
to estimate the pressure term, and also see \cite{Power2012} for a more detailed discussion about the virial ratio of halos. 
After applying the criteria described above, we have found that $\sim 21\%$ of the \coco{} halos and $\sim 13\%$ of the \scolor{} halos 
(for all mass range)
are not relaxed and were removed from further analysis.

\begin{figure}
 \includegraphics[angle=-90,width=0.48\textwidth]{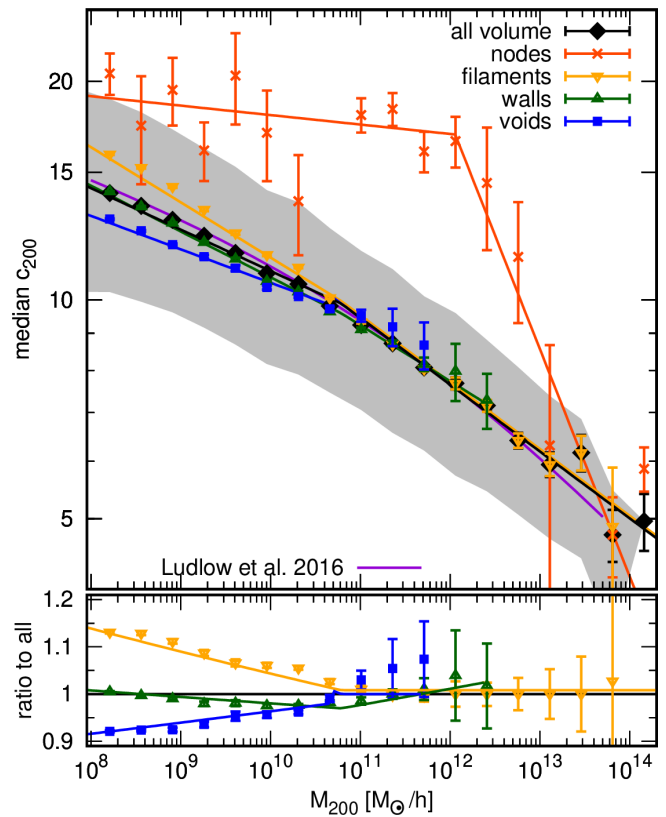}
    \caption{\coco+\scolor{} concentration-mass relation split across the different Cosmic Web environments. {\it The Upper Panel:} The $c(M_{200})$ relations for our joint samples. The data points and error-bars represent the median and its bootstrap error from the simulation data. The solid lines with matching color are the best-fitting broken power-laws (see the main text). The shaded region illustrates the 16-th:84-th percentiles for the all-volume sample. The purple line shows the prediction of Ludlow {\it et al.} 2016 model. {\it The Lower Panel:} The relative ratio of the data and the fits for void, walls and filament halos taken with respect to the all-volume sample.}
    \label{fig:c-M}
\end{figure}

In FIG.~\ref{fig:c-M} we show the median concentration-mass relation, $c_{200}(M_{200})$, for \coco+\scolor{} halos divided into 
different samples.
We consider {\bf all-volume} (black line), {\bf nodes} halos (orange crosses and lines), {\bf filaments} halos 
(yellow down-triangles and lines),
{\bf walls} halos (green up-triangles and lines) and halo found in {\bf voids} (blue squares and lines). For a comparison 
we show also the prediction
of Ref.\citep{Ludlow2016} model as a purple solid line. The error bars reflect the uncertainty with which we can calculate the median in each bin, which was obtained using 100 
bootstrap samples, while the shaded region illustrates the halo-to-halo spread, which is the 16-th and 86-th percentile of the 
halo population in each mass bin (it corresponds to the $1\sigma$ variance
for a normally distributed random variable). 

We notice that the modeled $c(M)$ relation of Ref.\citep{Ludlow2016} describes well the trend exhibit by the 
main all-volume sample. The things become interesting when we study the behavior of the samples belonging to different Cosmic Web segments. 
Here, it seems that the nodes-halos are described by starkly higher median concentrations than the overall population, and, moreover, for objects with
$M_{200}\simlt 10^{12}\Msun$ the $c(M)$ relation appears to be much flatter than what we can observe for the other samples. The remaining LSS environments
follow closely the main sample trend, with some appreciable scatter. This, however, applies only to halos more massive than $M_{200}\sim 10^{11}\Msun$.
For lower-mass objects the median concentrations in each web environment start to deviate systematically from the all-volume median.
Here, the void halos appear to be characterized by density profiles with lower concentrations than the universal sample, while 
the filaments halos, on the other hand,
have higher concentrations. Deviations of both samples seem to increase with decreasing halo mass. Interestingly, the wall sample is characterized by
the concentration-mass relation that is very close to the universal sample. To better understand and quantify the trends with halo mass and environment, we fit the $c(M)$ relation for
each environment with a broken power-law of the form:
\be
\label{eqn:broken-pl}
c_{200}(M_{200})=
\begin{cases}
 A (M_{200}/M_{th})^{b_{l}} & \text{for}\,\, M_{200}\leq M_{th}\\
 A (M_{200}/M_{th})^{b_{h}} & \text{for}\,\, M_{200}> M_{th}
\end{cases}
\ee

\begin{table}
\caption{The best-fitting $c_{200}(M_{200})$ parameters to the double power-law of Eqn.~(\ref{eqn:broken-pl}) for all the data 
samples shown in FIG.~\ref{fig:c-M}.
$M_{th}$ is expressed in $\times 10^{10}\Msun$, while max mass is in $\Msun$.
}
\large
\begin{tabular}{@{}|l|c|c|c|c|c|}
\hline\hline
Sample & $A$  & $M_{th}$ & $b_l$ & $b_h$ & max mass \\
\hline\hline
All-volume & 9.89 & 6.12 & -0.057 & -0.092 & $10^{14}$\\
\hline
nodes & 16.9 & 114 & -0.013 & -0.313 & $10^{14}$\\
\hline
filaments & 9.97 & 6.12 & -0.076 & -0.092 & $10^{14}$\\
\hline
walls & 9.59 & 6.12 & -0.063 & -0.077 & $3\times10^{12}$\\
\hline
voids & 9.72 & 6.12 & -0.046 & -0.092 & $6\times10^{11}$\\
\hline\hline
\end{tabular} 
\label{tab:cm_fit_params}
\end{table}

The single power-laws were used to described various $c(M)$ and has been shown to characterize reasonably well halos from various
simulations, albeit only for a limited mass range \citep{Neto2007,CMGao2008,Ludlow2010,Klypin2011,Munoz2011,Prada2012,Sanchez-Conde2014}.
We use double power-law for our samples, since it appears that
there is a close to universal mass-scale at which the various Cosmic Web populations deviate from the mean trend describing 
the overall halo population.
In Eqn.~(\ref{eqn:broken-pl}) the mass scale at which the broken-law changes the slope is defined as $M_{th}$, while $A$ is
the normalization
factor, $b_l$ and $b_h$ are the (l)ow and (h)igh-mass power-law slopes. We have found that the double power-law of the above 
form offers very good fits to all our data-samples. We have collected parameters for all the samples in Tab.~\ref{tab:cm_fit_params}. 
All fits have the same minimal mass of applicability set to $10^{8}\Msun$, the corresponding maximal masses are given in the table.

For the void sample, we could not robustly determine $b_h$, i.e. the power-law slope for the high-mass regime, since there are only a handful of halos with $M_{200}>M_{th}$ found in voids. Because
of this the variance of halo concentrations in that mass regime is considerably high, and in turn the whole void halo population can be well
described by a single power-law. Nonetheless, the void-sample still can be described very well by a double power-law, by using the $b_h$ value measured for the all-volume sample. The analysis of the best-fit parameters that describe our halo samples
highlights a couple of very interesting features. 

Firstly, we find that the population of halos found in node environments is very different
from the rest of the sample. Up to threshold mass of $M^{node}_{th}=1.14\times 10^{12}\Msun$ their concentration-mass relation is relatively flat,
slowly dropping with halo mass, with the power-law slope of $b_l=-0.13$. Above this mass, the node sample experience a dramatic shift to a steeply
declining power-law with slope of $b_h=-0.313$. This can be understood when we realize that the majority of small-mass halos found in nodes are
actually satellite halos found usually just outside the viral radii of much more massive halos found at the nodes of the Cosmic Web. 
Potentially many of them may be the so-called backsplashed halos, which 
traversed one or more times through the virial radius of a larger halo \citep{Behroozi2014,Adhikari2014,Busch2017,Diemer2020}.
Therefore, these low-mass objects,
although officially classified as distinct halos, can be to some extent subjects to preprocessing such as: tidal truncation,
stirring and disruption, just as regular subhalos found within the virial radius of their host halos.
Above the $M_{th}$ the node sample starts to be dominated by regular halos. This is also indicated by the growing fraction of halos 
found in nodes in that mass range (see again Fig.~\ref{fig:HMF}). It explains why the $c(M)$ experiences such a steep decline to 
arrive again at the universal all-volume median for $M_{200}\simgt 10^{13}\Msun$.

The three remaining Cosmic Web elements paint a more coherent picture. Here, all three samples approach the all-volume universal mean for the same
threshold mass of $M_{th}=6.12\times10^{10}\Msun$. Although, we find slightly different best fitting values for $b_h$, i.e. the high-mass slope of the $c(M)$ relation, the relation can be reasonably well described also by the all-volume high-mass fit. This specific threshold mass was surprisingly universal
across the different Cosmic Web segments and what is even more important, we find it to also has the same value when analysing the \coco{} 
and \scolor{} samples separately. Below the threshold mass we observe clear departure of the distinct samples from the all-volume median. 
Here, the void halos density profiles get less
and less concentrated compared to the universal mean, while the filament halos are characterized by denser central density profiles. 
Strikingly, the wall halo population appears to have the median concentration parameter that most closely follows the all-volume sample. 
This would indicate, that the wall halos find themselves in a perfect balance between higher-density filaments and empty voids to arrive 
at a median that is very close to the overall value. The maximum deviation between different populations is observed 
at $M_{200}=10^8\Msun$, where the median $c_{200}$ of the filament sample is higher by $\sim 15\%$ then the overall population.
Here the median void halo concentration is $\sim 8\%$ lower compared to the all-volume sample. For  halos with one order of magnitude higher masses these discrepancies
shrink to $8$ and $6\%$ respectively. Finally at masses $\simgt 10^{10}\Msun$ the three Cosmic Web samples start to quickly converge
towards the overall trend.

Our data do not allow to accurately study profiles below a halo mass of $\sim10^{8}\Msun$, so we are unable to check if the environmental trends
would continue. However, there is a hint in our data, seen both for voids and filaments, that the deviation from the all-volume median
starts to flatten at masses of $\sim4\times10^{8}\Msun$. This can be seen for the two least massive bins in FIG.~\ref{fig:c-M}. If this would be indeed the case,
then a more natural scenario would be favoured, where the environmental effects on the halo concentrations are, at best, saturating at around
our maximal deviations, rather than growing further with decreasing mass. Thus, we need to caution against extrapolating the trends
exhibited by our best-fits to lower halo masses.

Recently, in a work dedicated to study halo concentration over 20 decades in mass,
\citep{Wang2020} claimed that the halo concentrations are insensitive to {\it ``the local halo environment''}. Our results would then seem to be
in conflict with theirs. However, we note that the Ref.\citep{Wang2020} use a rather specific definition of a halo environment. First, they consider
only the local density, as measured in a sphere around a halo. Second, the density used for this specific proxy of a halo environment is measured
on a scale $5-10\times R_{200}$. Thus it is a scale-dependent density measure. In contrast, the \nexus+ environment is defined on scales 
that maximises the Hessian response signal, depicting and reflecting the multi-scale nature of the Cosmic Web \citep[see more in][]{Cautun2014}.
In fact, there is no correlation between our Cosmic Web flags and the local-density measure used by Ref.\citep{Wang2020}.
In addition, what plays here an important role is the fact that the Ref.\citep{Wang2020} used a series of nested zoom-in simulations,
which begins with a parent region of a relatively low density. Thus, starting from their level-2 (L2) zoom their halo population,
according to the definitions and criteria used in this work, would belong to only one specific \nexus+ environment. Since they place
each nested consecutive zoom-in region far away from massive halos, would most likely favour \nexus+ wall or void environments.
It would be interesting to see, if the environmental trends we have found hold down to lower halo masses. Such a study
will require a separated dedicated set of zoom-in simulations, similar to the ones employed by Ref.\citep{Wang2020} and 
we leave this aspect for future work.

Whether the mass threshold of $M_{th}=6.12\times10^{10}\Msun$ is a universal parameter related to the $\lcdm{}$ Universe and the NEXUS+ method 
is a subject of an open debate. We can suspect that this new mass scale might be a function of the background cosmology, but 
other factors like the impact of a simulation volume or mass resolution cannot be excluded at this time. 
A more detailed study that would involve 
usage of many different N-body simulations would be required to address this question. We leave such a study for the future. 
However, the implications
of the existence of $M_{th}$ below which the universality of the concentration-mass relation is broken, might be profound. 
Especially, when we recall
that below this mass scale the fraction of halos inhabiting each of the three different environments is significant and 
approach an equal division.
We will discuss the consequences of this result in the concluding section.

\subsection{Mass assembly histories}
\label{subsec:MAHs}
\begin{figure}
    \includegraphics[angle=-90,width=0.48\textwidth]{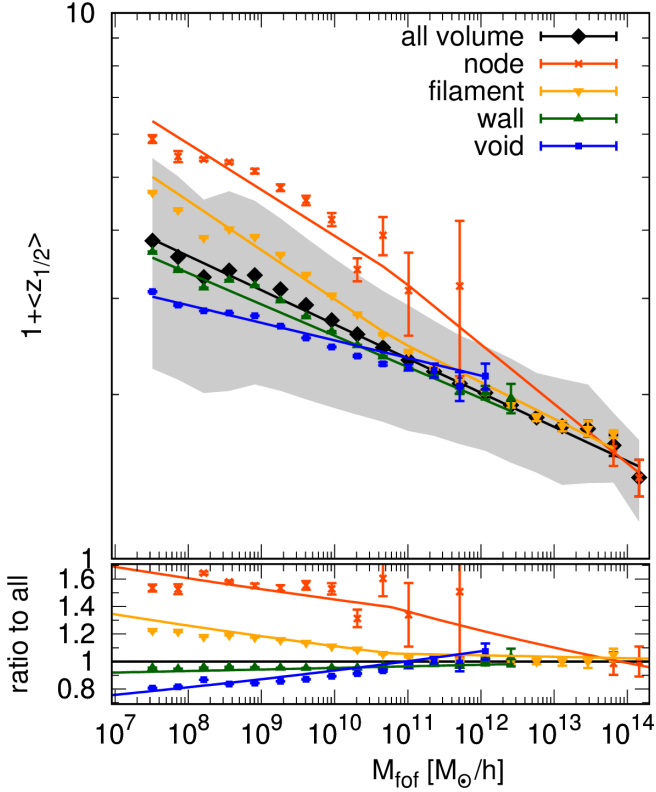}
    \caption{ The average \coco+\scolor{} formation redshifts, $\langle z_{1/2}\rangle$, for halos split according to their web environment. {\it The Upper Panel:} The halo formation redshifts plotted versus $z=0$ FOF halo mass. The points with error bars correspond to our simulation data, while the solid lines are the best fitting relations (see the main text). The shaded region marks the 68 percentiles for the all-volume sample. The error bars indicate bootstrap errors on the mean. {\it The Lower Panel:} The ratios of the formation redshift in the different environments with respect to the all-volume sample.
    }
    \label{fig:M-z}
\end{figure}

In the previous section we have found a clear effect of the different Cosmic Web environments on the median halo 
concentrations. In hierarchical structure formation cosmologies, such as the CDM model, halo concentrations 
are correlated with some
characteristic halo formation times. This reflect the fact that due the hierarchical bottom-up build-up of the halos the inner-most 
regions of dark matter halos are the first one to be assembled. The material acreeted later has usually higher angular momenta, which means 
that it seldom is able
to settle down in the central parts of the halo mass distribution. In such a picture the inner parts of the halo density profiles 
are dynamically old, and the density reached there during the assembly time sets the halo concentration. Since the averaged background 
density of the Universe drops with time, older halos have usually higher central density and also higher concentrations. 
For the $\lcdm${} model this scenario is evident in a wealth of simulations-based studies \citep{Ludlow2010,Ludlow2016,Rey2019,WangKuan2020,Chen2020}.

The clear correlation between the redshift of the halo assembly (also called the formation redshift) and the halo concentration suggest that
we should also see the effect induced by different Cosmic Web environments in the halo assembly histories. For each halo, we define the formation redshift, $z_{1/2}$, as the redshift at which 
the most massive progenitor (MMP) of the halo reaches half of the final $z=0$ halo mass (\citep[see][for other possible definitions]{Li2008}).
Thus, we  define
\be
\label{eqn:zfrom_def}
M(z_{1/2})\equiv \frac{1}{2} M(z=0)\,.
\ee

To find $z_{1/2}$ for the \coco{} and \scolor{} halos we follow the halo merger trees constructed as described in the section \S\ref{sec:simulation}.
Here, the mass of the MMP branch at each redshift is that of it's parent FOF halo at that time. Next, we bin halos according to present-day $M_{FOF}$ and calculate the mean formation redshift, $\langle z_{1/2}\rangle$, in each bin and for each web environment. 
The obtained
$z_{1/2}(M_{fof})$ relation is described by a linear fit in $\log(1+z_{1/2})$, same as in \cite{MS2,COCO1},
which takes the form:
\be
\label{eqn:m-z-fit}
1+\langle z_{1/2}\rangle = A_{half}\left({M_{FOF}\over M_{th}}\right)^{\beta_{half; \; 1}}\,.
\ee
In previous studies, this power law was rescalled by a $10^{10}\Msun$ mass, however here we choose 
to use the characteristic environment threshold mass of 
$M_{th}=6.12\times 10^{10}\Msun$ found in the previous paragraph. 
Such a single power-law turned to be a good fit to the all-volume,
wall and voids samples. For the filaments and nodes samples we had to use a two-regime power law
\be
\label{eqn:m-z-fila}
 1+\langle z_{1/2}\rangle =
\begin{cases}
 A_{half}\left({M_{FOF}\over M_{th}}\right)^{\beta_{half; \; 1}} & \text{for}\,\, M_{FOF}\leq M_{th} \\
 A_{half}\left({M_{FOF}\over M_{th}}\right)^{\beta_{half; \; 2}} & \text{for}\,\, M_{FOF}> M_{th}\\
\end{cases}
\ee
with two power-law breaking mass scales set to be at the environment
threshold mass $M_{th}$. 
We give the best-fit parameters of the averaged  $1+\langle z_{1/2}\rangle$ relations for all our data samples in 
Tab.~\ref{tab:m-z-table}.

\begin{table}
\caption{The best-fitting parameters for the dependence of the halo formation redshift on halo mass. 
The dependence is given by the power-laws of Eqn.~(\ref{eqn:m-z-fit})-(\ref{eqn:m-z-fila}) and are shown in FIG.~\ref{fig:M-z}. 
The rows that do not have a value for $\beta_{form; \; 2}$ mean that they were well fit 
by the single power-law given in Eqn.~(\ref{eqn:m-z-fit}).}
\large
\begin{tabular}{@{}|l|c|c|c|c|}
\hline\hline
Sample name & $A_{form}$  & $\beta_{form; \; 1}$  & $\beta_{form; \; 2}$\\
\hline\hline
All-volume & 2.40 & -0.063 & --\\
\hline
nodes & 3.35 & -0.084 & -0.109 \\
\hline
filaments low & 2.54 & -0.09 & -0.067\\
\hline
walls & 2.31& -0.057 & --\\
\hline
voids & 2.37& -0.032 & --\\
\hline\hline
\end{tabular} 
\label{tab:m-z-table}
\end{table}

In FIG.~\ref{fig:M-z} we plot the average formation redshifts, $1+\langle z_{1/2}\rangle$, and their associated best fits (the Upper Panel) for all halos and for halos in each web environment. To better highlight the trends, the bottom panel of  the figure shows the relative difference with respect to the all-volume sample for our joined \coco+\scolor{} data. Despite the fact that some of 
the CW components (\ie{} the nodes and filaments halos) experience a more complicated multi-slope relation, we clearly 
identify a common feature to the all data samples: a monotonic decline of the formation redshift with increasing 
halo mass. This reassures that in all the CW elements the halo build-up is still, as expected, progressing in a hierarchical
manner. Comparing the different environments, we see that below the characteristic threshold mass $M_{th}=6.12\times10^{10}\Msun$ 
the formation redshift at fixed halo mass depends on the CW component in which a halo is located. 
The voids halos have significantly lower $z_{1/2}$ compared to the all-sample mean, while on the other
hand the filament population have higher formation time mirrored in a nearly prefect way to the void sample. 
The node FOF groups are the oldest in our runs, thus the NEXUS-node sample trace the first halos to form in the Universe. 
Here, the wall population has a value a bit lower than the assembly times for the whole population, however, the effect is small and contained to 
within $5\%$. 

Similarly to the density profiles, the trend with environment becomes more and more prominent as we decrease the halo mass. 
What is also really noticeable is that the trend with the CW is at the same level as that seen for $c_{200}$, reaching a maximum $20\%$ difference for filament and void halos at $M_{FOF}=3\times10^{7}\Msun$.
It highlights how the local CW environment moderates the merger and accretion rates of halos 
that, in turn, is reflected in the halo formation times. The effect is only significant 
for halos less massive than a characteristic environmental mass scale, $M_{th}$.

\subsection{The $\vmax$-$\rmax$ relations}
\label{subsec:vmax-rmax}
\begin{figure}
 \includegraphics[angle=-90,width=0.48\textwidth]{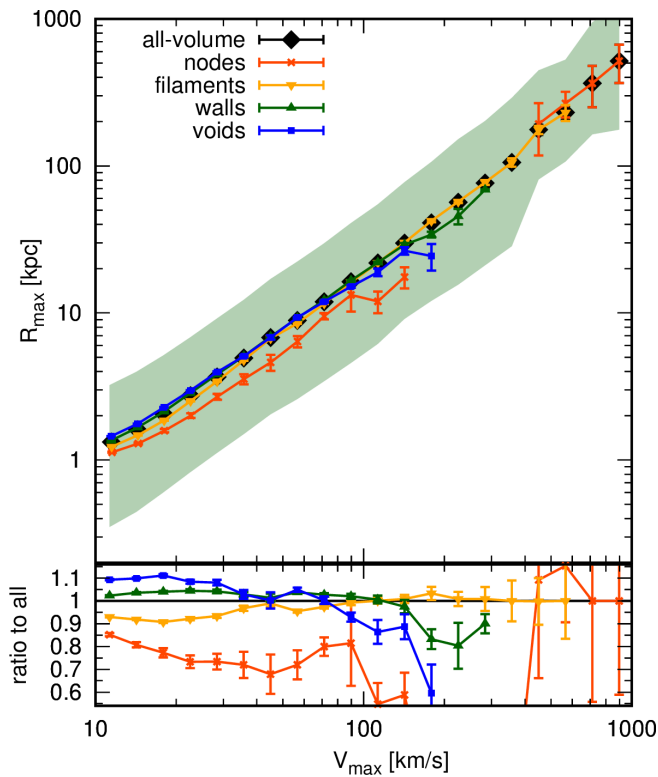}
 \caption{The median \coco+\scolor{} $\rmax-\vmax$ relation for halos in different Cosmic Web structures.
 {\it The Upper Panel:} $\rmax-\vmax$ for all samples. The errors on the data points reflect the bootstrap errors on the medians.
 The shaded green region marks the 15-th to 84-th percentile spread around the all-volume sample. 
 {\it The Lower Panel:} The relative ratio of each Cosmic Web population to the all-volume sample.}
\label{fig:vmax-rmax}
\end{figure}

Another complementary probe of the halo mass distribution is the shape of the circular velocity curve,
which due to its cumulative nature is less prone to small
density fluctuations due to, \eg{} substructures, as is the case for the differential measure of $\rho(r)$. 
The circular velocity is defined as:
\be
\label{eqn:circ_vel}
V_{c}(r) = \sqrt{{\textrm{G}M(<r)\over r}}\,,
\ee
where $M(<r)$ is the mass enclosed inside a sphere of radius $r$ centred at the halo centre. 
For a halo that exhibits perfect spherical symmetry, 
the circular velocity, $V_c(r)$, is exactly equal to the circular orbital velocity at distance $r$. 
For well resolved and relaxed halos, the circular velocity takes only one maximum value, $\vmax$, that occurs at 
a radial distance, $\rmax$.
Now, studying the $\vmax-\rmax$ relation gives a way to characterize the halo internal mass profile, based on 
its internal dynamics. The $M(<r)$ profile is simply a linear convolution of the halo density profile.
Thus in principle, the information contained by the $\vmax-\rmax$ is the same as the one
encoded in halo density profiles. However, the dynamical measure of $\vmax$ is, in principle, easier to access
by observations than the measurement of the halo concentration profile,
which is why we perform a separate analysis of the $\vmax-\rmax$ relation.

Here we investigate the $\vmax-\rmax$ relation for our joint main halo samples. Generally we can expect that 
the smaller $\rmax$ at fixed $\vmax$ the steeper the inner $M(<R)$ profile. Indicating a halo with a denser inner
region and by logic a higher concentration parameter. However, due to the cumulative nature of 
the $M(<R)$ distribution the $\vmax-\rmax$ relation 
is more stable against shot-noise. We will use this property to check and probe the environmental effects down to halos even 
smaller (\ie{} $\vmax= 10\kms$) than this was possible for $\rho(r)$. 

We start by constructing the joined \scolor+\coco{} sample, this time we consider all \coco~ halos down to $\vmax=10\kms$,
which is the convergence limit for $\vmax$ for this run (see \cite{COCO1} for more detailed analysis). 
The \scolor~ population is susceptible to resolution effects at higher
velocity values due to combination of lower mass, but also force resolution. Thus we set a lower-cut for \scolor~ halos at $\vmax=40\kms$, 
and consider only halos above this threshold for our composite sample. Lastly, we use the Springel \etal{} \citep{Springel2006} approach to correct the measured $\vmax$ values for effects arising due to the force softening used in N-body gravity calculations, which consists 
of lowering the maximum velocity, $\vmax$, of low-mass objects whose $\rmax$ is comparable to the gravitational 
force softening of the simulation. We apply the correction formula proposed by \citep[][Eqn. (10) therein]{Aquarius}, 
which, under assumed perfect circular orbits, accounts for this effect. After applying all the above steps we obtain a sample of $\vmax-\rmax$ pairs that robustly sample the maximum of circular velocity 
over three orders of magnitude, \ie{} $10\leq \vmax/(\kms)\leq 1000$. The corresponding halo mass range then is from 
$2\times10^7\Msun$ to $10^{14}\Msun$ (see \eg Fig.~A1 in \cite{COCO1}).

In FIG.~\ref{fig:vmax-rmax} we show the median $\vmax-\rmax$ relation for all our halo populations.The shaded green 
region illustrates the spread  around the all-volume sample as measured by the 16-th and 84-th percentiles, 
which would correspond to $1\sigma$ dispersion for a Gaussian-distributed random variable. 
First, we find that the the spread has a nearly constant width
(as measured using relative ratio) for the whole range of probed maximum circular velocities. Secondly, we see 
that region with $\vmax\simlt 15\kms$ seems to be already affected to some extent by the resolution effects, which is indicated 
by the small, albeit noticeable, flattening of the relation. We opt to keep this data in the comparison, as we expect that
this resolution effects would affect alike all Cosmic Web components. Henceforth the environmental 
signal encoded there is still useful, provided that for $\vmax\simlt 15\kms$ we study only the relative ratios. 

A number of interesting features visible in figure \ref{fig:vmax-rmax} deserve further
attention. The effect of the Cosmic Web environment is also clearly visible here, and what is very reassuring 
the magnitude of $\rmax$ reduction (boost) is consistent with the measured increase (reduction) of halo density 
concentrations seen in FIG.~\ref{fig:c-M}. This again, as we already seen it before, follows a nearly mirrored effects 
for voids and filament halos, reaching a maximum effect of $\sim11\%$ for both environments at $\vmax=18\kms$, 
which corresponds to halo mass of $\sim4\times 10^{8}\Msun$. The net-effect of the environmentally-driven $\vmax$ 
boost or reduction is a few percent-points smaller than what we observed for median $c_{200}$, however this is not 
surprising taking into account the fact that $V(R)$ is cumulative. What is very important is that using $\vmax-\rmax$ data
we can now confirm the flattening (or saturation) of the Cosmic Web induced effect appearing for 
$M\simlt 4\times 10^8\Msun (\vmax\simlt20\kms$).
Which was previously only hinted by our $c_{200}-M_{200}$ data. This result indicates that our fits for 
the environmental deviations from the universal $c(M)$ relation shown in Tab.~\ref{tab:cm_fit_params} 
by no means should be extrapolated below the minimum $10^{8}\Msun$ mass. A more realistic modeling would 
seem to consists of a saturation of the difference, rather then the extrapolation of the trend seen at higher masses. 

Another feature seen in our data is that, if we disregard the discrepancies between voids and walls samples for large
objects (\ie $\vmax\geq100\kms$) where the uncertainty is already high, we observe that the void, filament, and 
wall halos converge to all-volume value for $\vmax\geq 80\kms$. This value of the maximum of the circular velocity 
curve corresponds to halo mass of $5-6\times10^{10}\Msun$, a value very close to the universal environmental 
threshold mass, $M_{th}=6.12 \times 10^{10}\Msun$, we found for the $c(M)$ and $z_{1/2}(M)$ relations. 
For the node halos the $\vmax-\rmax$ relation reveals a picture that is very consistent to what we have seen before. 
Only big and massive halos (in the galaxy group and cluster mass regime) follow the all-volume trend  -- this 
is expected by construction, since in this mass range the majority of halos live in the node environment. 
At lower masses, in the whole probed regime of $10\leq \vmax$(s/km)$\leq 150\kms$ the node halos 
have smaller $\rmax$ values at fixed $\vmax$, which corresponds
to more compact and concentrated density profiles. The reduction in $\vmax$ is substantial and typically
is of the order of $20$ to $30\%$ percent, when compared to the all-volume sample. The visible trend of 
increasing $\rmax$ reduction when going from smaller to larger $\vmax$ values is fully consistent with 
the relative flatness of $c(M)$ for nodes, since larger objects would have typically smaller concentrations, 
thus to obtain a nearly flat trend the decrease in $\rmax$ needs to be larger.

\subsection{Shape and spin}
\label{subsec:shape_spin}

So far we focused on the one primary halo internal property, namely its mass distribution. There are of course more 
properties intrinsic for each DM halo, that are of importance and interest. The other two that are commonly 
measured in simulations and used in a variety of modelling are the shape and spin parameters. The former is 
related to the symmetry of the internal mass distribution as traced by the halo mass inertia tensor, while 
the later is used as a measure of the net bulk internal rotation and is usually characterized using the halo angular momentum.
\begin{figure}
 \includegraphics[angle=-90,width=0.48\textwidth]{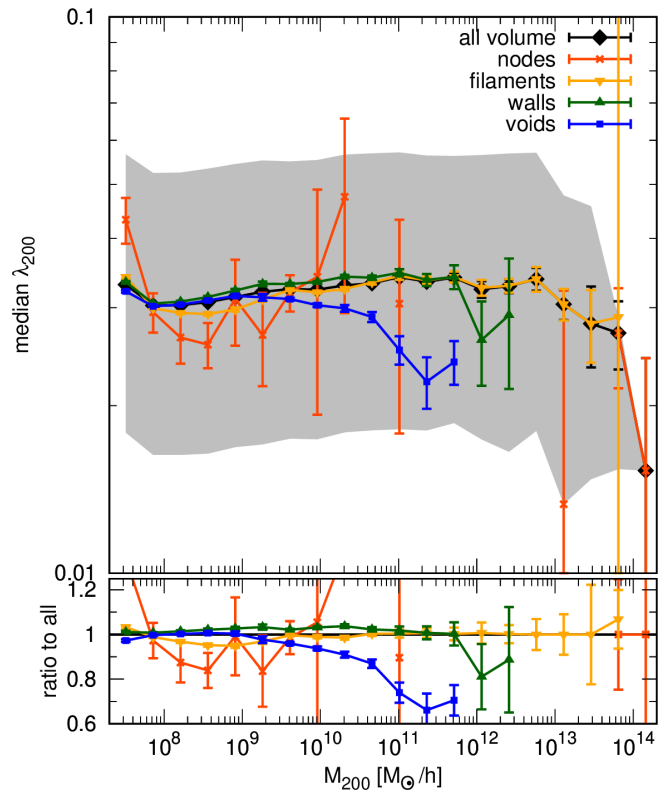}
 \caption{The median \coco+\scolor{} halo spin, $\lambda_{200}$, as a function of halo virial mass, $M_{200}$, 
 and Cosmic Web environment.
 {\it The Upper Panel:} $\lambda_{200}(M_{200})$ relation for all samples. The errors on the data points 
 reflect the bootstrap errors associated to the median calculation.
 The shaded gray region marks the 16-th:84-th percentile spread of the all-volume sample. 
 {\it The Lower Panel:} The relative ratio for each Cosmic Web halo population with respect to the all-volume sample.}
\label{fig:j200-m200}
\end{figure}

We first take a look at the halo degree of rotational support as measured by the halo spin parameter, 
$\lambda$. To calculate it, we use the Bullock
definition \citep{Bullock2001} 
\be
\label{eqn:spind_def}
\lambda = {|j|\over {\sqrt{2}R_{200}V_{200}}}\,\,,
\ee
where $V_{200}$ is the value of the circular orbit velocity from Eqn.~(\ref{eqn:circ_vel}) taken at 
the virial radius $R_{200}$, and
$j$ is the specific angular momentum of the halo
\be
\label{eqn:specific_j}
j={1\over N_p}\sum^{N_p}_{i} r_i\times v_i\,.
\ee
There are other definitions of this parameter (\eg \citep{Peebles1969}), but the formulation from Ref.\citep{Bullock2001} 
is one of the most convenient to measure from simulations. The spin parameter characterizes to 
what extent the halo is rotational supported. Halos with low spin are 
dominated by velocity dispersion of random motions and have low degree of overall rotation. On the other hand, 
we can expect that halos with high $\lambda$-s show more signs of coherent rotation in their orbital structure. 
The $\lambda$ parameter for $\lcdm{}$ cosmology has been studied in a number of previous works 
\citep[\eg][]{Steinmetz1995,Cole1996,Maccio2007,Hellwing2013,Braun-Bates2017},
and it has been generally found that
DM halos are characterized by low spin values, consistent with a marginal degree of a coherent rotation. 
The distribution of the spin parameter
has been found to be log-normal with the mean (\ie{} the first moment) 
between $0.04-0.05$ at $z=0$ \citep[\eg][]{Warren1992,Cole1996,Vitvitska2002,Bett2007}. These log-normal mean values 
corresponds to
median values in the $0.02\simlt \lambda_{med}\simlt 0.04$ range.
In FIG.~\ref{fig:j200-m200} we show using our usual pattern of two panels the median virial spin parameter 
as measured at $R_{200}$ for halo samples split among different Cosmic Web elements. 

The environmental effects in the spin parameter that we can see in the Figure follow much less obvious 
patterns than it was the case for the density profiles. Noticeably, here the effect of the environment 
appears to be present only for more massive halos, rather than for the lowest-mass objects like we have found in the previous section.
The node-halo sample again has the largest uncertainties in the median value, and with our limited statistic
it is hard to draw any firm conclusions. On the other hand, the effect seen in the void population is quite
prominent and very interesting. Starting from $M_{200}\geq 3\times10^{10}\Msun$ that sample experiences 
a growing departure from the universal all-volume trend. At $2-4\times10^{11}\Msun$, which is the high-mass end of the void population, 
the reduction in $\av{\lambda_{200}}$ 
reaches nearly $\sim35\%$, which is quite a strong spin reduction. The most massive end of the wall population shows 
a hint of a similar trend, alas our statistics are too poor to confirm this in a robust way. Interestingly, 
for $M_{200}\leq5\times10^{11}\Msun$ the wall sample shows
very small, yet consistent and significant excess of the spin with respect to the main sample. In contrast, 
the median spin of the filament halos of masses $10^8\leq M_{200}/(\Msun)\leq 4\times 10^9$ is smaller by $\sim 5\%$ then the main sample.

There are different ways in which halos acquire non-vanishing total angular momentum. In the linear 
and weakly non-linear regime the spin is generated by torques induced by tidal fields associated with the local large-scale structure. 
This idea was first formulated as the Tidal-Torque-Theory (TTT) to explain the angular momentum of galaxies 
\citep{Peebles1969,White1984,Barnes1987,Heavens1988,Catelan1996}.
The TTT explains well the growth of the angular momentum
at early times. The insight from N-body simulation indicated 
that once the halo experience more rapid merger rates
and mass accretion rate the primordial spin acquired via tidal torques becomes subdominant. 
This is because the specific angular momentum
$j$ from Eqn. (\ref{eqn:specific_j}) is effectively a mass weighted quantity, as we use N-body 
particles as tracers. 
At later times the majority of halo particles consists of the material that resides in more 
outer-parts and their net angular momentum tends to be 
a bit higher reflecting both the increased radial distance and 
the non-linear character of halo mergers and late-stage of mass accretion
\citep{Porciani2002A,Porciani2002B,Robertson2006,Lopez2019}. In this picture,
halos that would live in a much less violent and crowded environment would naturally express 
a lower  spin. This is exactly
what our results for voids and walls populations indicate. A bit puzzling is what we 
observe for low mass node halos, which tend to
have marginally significant lower spin then the all-volume sample. This merit a more thorough 
investigation, and we lave such for future work.

\begin{figure}
 \includegraphics[angle=-90,width=0.48\textwidth]{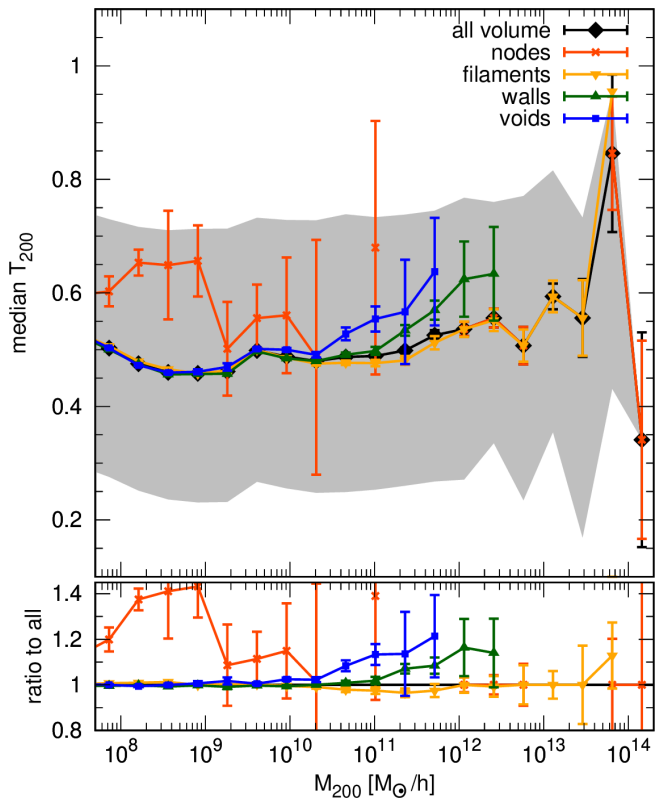}
 \caption{The \coco+\scolor{} median halo triaxiality shape parameter, $T_{200}$, as a function 
 of halo mass, $M_{200}$.
 {\it The Upper Panel:} $T_{200}(M_{200})$ relation for all samples. The errors on the data points 
 reflect the bootstrap errors around the medians.
 The shaded gray region marks the 16-th-84-th percentile spread around the all-volume sample. 
 {\it The Lower Panel:} The relative ratio of each Cosmic Web population to the all-volume sample.}
\label{fig:T200-m200}
\end{figure}
To measure the halo shape we use the eigenvalues of the halo's mass tensor of all the
particles within $\leq R_{vir}$ that do not belong to any substructure:
\be
\label{eqn:inertia_tensor}
I_{ij}={{1}\over{N_p}}\sum_{N_p}x_ix_j\,\,,
\ee
where the particle positions $x_i$ and $x_j$ are with respect to the center of mass of the halo and the sum
is over all the particles that belong to a given halo. The eigenvalues of this tensor corresponds to squares of 
the principal axes of the halo shape ellipsoid. We sort the eigenvalues and normalize them by the largest axis:
\be
\label{eqn:shape_axes}
a>b>c\,, \quad \tilde{b}=b/a\,,\quad\tilde{c}=c/a\,,\quad \tilde{a}=1\,.
\ee
In general, we expect halos to be tri-axial ellipsoids that can be either prolate or oblate 
\citep[\eg][]{Warren1992,Cole1996}.
The halo overall shape 
can be reasonably well described by a single parameter that is a combination of the all three eigenvalues.
This is the halo triaxiality parameter
\be
\label{eqn:trixality}
T_{200}\equiv {{a^2-b^2}\over{a^2-c^2}}= {{1-\tilde{b}^2}\over{1-\tilde{c}^2}}\,.
\ee
Values of $T$ closer to unity indicate a shape that is closer to a prolate ellipsoid, in contrast low 
$T$ value corresponds to an oblate halo.

We study the median triaxality as a function of halo mass in FIG.~\ref{fig:T200-m200}, where
two panels show 
the absolute values (the upper one) and the ratios with respect to the all-volume sample (the bottom panel). The gray 
shaded region shows the dispersion
measured by the 16 to 84-th percentile spread around the median all-volume sample and the symbols with error 
bars indicate the uncertainties associated to the median calculation. The general trend visible for all environments
except the nodes is consistent with
a common picture emerging from N-body simulations. Namely, that more massive halos tend to be more prolate.
This reflects their younger dynamical states compared to the less massive halos 
\citep{Allgood2006,Bonamigo2015,Despali2017,Lau2020}.
Focusing on specific environmental trends, the plots in the figure suggest that at small masses void, 
walls and filament samples
converge to the same value. In contrast, the cluster-environment halos are characterised by a significantly 
higher triaxality values, 
albeit again this sample suffers from the biggest uncertainties due to small number statistics. 

Starting from $\sim2\times 10^{10}\Msun$ for voids and $\sim 10^{11}\Msun$ for walls,
we observe an interesting trend of both samples manifesting an excess triaxality compared to the filaments 
and all-volume groups. 
The picture painted by the triaxiality-mass dependence complements very well what we have seen previously 
for the spin parameter. The samples
that has lower bulk rotation show also higher degree of triaxiality. Generally, 
a highly prolate shape is thought to arise shortly after the initial halo collapse,
which never happens simultaneously along all three major axes.
Therefore, the halos whose shapes were not significantly affected
by a recent merger event bear this mark of initial not-perfectly aligned collapse. In addition, 
the nearby tidal forces also inflict
some effect by distorting the halo shapes \citep{Porciani2002B,Maccio2007,Bett2007}.

The emerging picture here is the following. If we exclude the nodes sample then the halo shapes and 
spins are affected by the Cosmic Web 
environment only for halos more massive than a few $\times 10^{10}\Msun$. Recalling the results from 
FIG. ~\ref{fig:M-z}, halos in this mass regime have on average the same formation time and concentration as the mean sample. 
Thus, the increased triaxiality and reduced
spin observed for void and wall fractions is resulting from the interaction with the local tidal 
forces and thanks to much more quiescent 
environment of void and wall regions this interaction is not trampled by intensive late halo bombardment. 

\section{Conclusions}
\label{sec:conclusions}

In this paper we have analyzed halo populations from the {\it Copernicus Complexio} suite of high-resolution 
N-body simulations to find to what extent their properties are affected by large-scale environments.
We define and categorize the halo environment in terms of the four distinct 
Cosmic Web elements: nodes, filaments, walls and voids. These are identified by 
applying the \nexus+ algorithm to the \scolor{} simulation box.
    
Using the very high resolution of \coco+\scolor{} run we were able to analyze mass functions and internal 
halo populations for halos spanning 6 decades in mass. Thus, for the first time we study the effects 
of the Cosmic Web on halos with masses from $\sim10^{8}\Msun$ to $\sim 10^{14}\Msun$. 
Our results concerning shape, density profile, and spin for massive halos are in good agreement with the trends 
found by earlier studies that have used the \nexus+ classification algorithm \cite{Cautun2013,Cautun2014}.
However, in the regime of $M\leq 6\times 10^{10}\Msun$, which has not been widely explored by previous studies, we find a number of new and interesting features. 
Here we list again and comment on the most important findings.\\\\
{\it Large-scale Cosmic Web:}\\
  We get similar results for the mass and volume filling fractions as in the original \nexus+ paper,  finding that the most of 
  the Universe volume belongs to voids ($78.8\%$), but at the same time they contain only $13.7\%$ of mass. This amounts to an average
  density of $\rho_v=0.17\times\Omega_m\rho_c$. A fraction of $16.8\%$   of the volume and $21\%$ of mass 
  is found in walls, which corresponds to an average density contrast,  
  $\rho_w=1.24\times\Omega_m\rho_c$. More than half ($52\%$) of the mass in the Universe resides inside filaments, but
  they only take $4.2\%$ of the volume. This makes the filaments already quite dense, with an averaged density roughly $12.4$-times higher 
  then the background mean density. The node or cluster-like dense environments are very rare, 
  spanning only $0.045\%$ fraction of the volume. This minute fraction however contains as much as $13.4\%$ of the total mass. 
  Thus, confirming that the \nexus+ nodes are very compact regions with an averaged density contrast of $\av{\de_c}=299$.\\\\
{\it Halo Mass Function:}\\
We studied the HMF using a joint \coco+\scolor{} sample of halos, where the the 
\scolor~ sample was supplemented at the low-mass end, i.e. $M\leq3\times10^{8}\Msun$, with the \coco~ halos. 
\begin{itemize}
 \item We find that in the $10^{12}-10^{13}\Msun$ mass range the vast majority of halos (\ie{}$\sim 95\%$) are found 
 in filaments.
 \item In the regime of $10^8-5\times10^9\Msun$ the ratio of halos found in walls and in filament is approximately 
 equal, with void fraction growing considerably with decreasing halo mass.
 \item At the low-end of the HMF resolved by \coco{}, $M_{FOF}\sim 10^{7}\Msun$, we find that the fraction of halos found in filaments, 
 walls, and voids is roughly equal to $1/3$ for each. 
 \item The void HMF has a decline starting at $M_{FOF}\simgt 10^{10}\Msun$, while the wall HMF experience 
 a sharper cut-off starting at $M_{FOF}\simgt 10^{12}\Msun$;
 \item In contrast the filament HMF first follows closely the wall-one, but above the mass close to the environmental
 threshold $M_{th}=6.12\times10^{10}\Msun$ starts to grow, only to be exponentially suppressed at 
 $M_{FOF}\simgt10^{13}\Msun$.
\end{itemize}
{\it Halo density profile:}\\
We find halo concentrations by fitting the NFW profile to well resolved halos (\ie{} with at least 5000 particles). 
Prior to this, we remove all unrelaxed halos. We find that:
\begin{itemize}
 \item The node sample is an outlier. It experiences median concentrations that are typically $\sim 50\%$ larger
 then the all-volume sample. The shape of $c(M)$ relation is also quite different from the rest of the environments.
 Concentration-mass dependence of node halos is flat for $M\simlt 10^{12}\Msun$, and becomes a very steep 
 power law for higher masses.
 \item The void, wall, and filament samples have a universal threshold mass of 
 $M_{th}=6.12\times10^{10}\Msun$. Above this mass, all halos have the same mean concentration indifferently of their host environment. 
 Below $M_{th}$ we observe an increasing trend of halo concentrations with web environment. 
 \item The low-mass halos in filaments have higher concentrations than the overall sample, with the difference being as high as 14\% for $M\simeq 2\times10^8\Msun$.
 \item In contrast, void halos have the lowest concentration; on average $\sim7\%$ lower than the overall population at $M\simeq 2\times10^8\Msun$.
\end{itemize}
{\it Mass Assembly Histories:}\\
We built merger trees at the subhalo level and followed the most massive progenitor to link FOF groups at different redshifts.
\begin{itemize}
 \item The dependence of halo formation times mirrors the trends seen for halo concentration, albeit with slightly larger net effects. At fixed halo mass, the oldest halos are found in nodes, followed by filaments, and the youngest are in voids. 
  \item The average formation redshift of the Cosmic Web segregated halos also can be characterized by the same universal mass threshold 
 of $M_{th}=6.12\times 10^{10}\Msun$; only for lower masses we find an environmental trend.
 \item The formation redshift--mass relation of void and wall halos are well described by single power-law. In contrast, filament and node halos show a more complex mass dependence and a double-slope power-law need to be used.
 \item The nodes halos have the steepest $\av{z_{1/2}}(M)$ relation and it indicates that this sample exhibits the most hierarchical buildup and contains the oldest halos in the Universe.
\end{itemize}
{\it Internal dynamics traced by the $\vmax-\rmax$ relation:}\\
This offers a complementary picture of halo density profiles and allows us to study considerably lower-mass halos, 
down to $\vmax\sim10\kms$ (this corresponds to a virial mass of $\sim10^7\Msun$).
\begin{itemize}
 \item At fixed $\vmax$, we find that node and filament halos have the lowest $\rmax$ values and voids the largest.
 \item The trend with environment is significant only for halos with $\vmax\simlt 80\kms$, which corresponds to a mass scale of
 ${\sim}6\times10^{10}\Msun$, very close to the same environmental mass threshold, $M_{th}$, we discussed previously. 
 \item We find that the trend with Cosmic Web environment flattens for low-mass halos, confirming what was previously 
 just hinted for when studying $c(M)$.
\end{itemize}
{\it Spin and shape parameters:}\\
We characterized the halo rotation using the dimensionless spin parameter, 
$\lambda$, and the halo shape using the eigenvalues of the mass tensor.

\begin{itemize}
 \item The general trends for the mass-environment effects appear to be reversed for both halo spin and triaxiality. 
 Here, the more massive rather than the low-mass halos show a trend with environment.
 \item For the spin, only the void sample showed any significant deviation from the universal mean, with a trend
 of reduced spin that starts at $M_{200}\simgt3\times10^{10}\Msun$. For halos with masses larger than $\sim10^{12}\Msun$,
 the effect seems to saturate at nearly $35\%$ spin reduction.
 \item Some hint of a similar effect at an order of magnitude higher mass is present for wall halos, with a
 spin reduction of $\sim20\%$.
 \item The triaxiality parameter of void and wall samples show and excess starting from $M_{200}=3\times10^{10}\Msun$(voids) and
 $10^{11}\Msun$ (walls). This indicate that large halos living in those two less dense environments tend to be
 more prolate.
 \end{itemize}

The results obtained here made use of the \subfind halo finder and the \nexus+ cosmic web identifier. We expect that using 
another halo finder would impact the results minimally, since for example \citep{Knebe2011} has shown that most halo finders agree to better 
than $10\%$ in terms of halo abundance and profiles; with the differences unlikely to be correlated to the local  environment of 
a halo. In contrast, there is a larger difference between the Cosmic Webs identified by the various finders used in literature\citep{Libeskind2018}.
This means that using another web finder might result in quantitatively different trends. Instead of being a limitation, this 
instead can be seen as an opportunity. By analysing how halo properties vary with environment for different web finders we 
can identify the method that maximizes the environmental trend, which is potentially the cosmic web definition that best 
captures the physical processes affecting halo assembly. This is similar to the approach taken by \citep{Cautun2018,Paillas2019,Davies2021}
who have compared which void finders are best for testing alternative cosmological models.

The picture emerging from our analysis highlights the important role that the Cosmic Web and more generally large-scale structure 
plays in nurturing the growth and evolution of dark matter halos. The magnitude and mass scales at
which some environmental
effects can be seen are varying with halo properties and Cosmic Web component. In general, it is quite clear from our analysis,
that the four Cosmic Web elements we consider: voids, walls, filaments and nodes, create unique ecosystems, 
each differing from the other by more than a mere measure of the local density. This is inline with previous findings
of Ref.\citep{AragonCalvo2010}, who emphasised that density alone, as a criterion for defining the Cosmic Web elements,
fails to capture important dynamical and connectivity aspects of the Web.

The trends with web environment that we have found in the median concentration-mass relation are likely connected
with the variation in halo assembly histories with environment, supporting also the well-known assembly-bias of DM halos.
In this context the model of $c(M)$ proposed by Ref.\citep{Ludlow2010,Ludlow2016} seems to be the most
physically motivated. However, our study indicates that the trends in the assembly times are not simply one-to-one 
translated to differences in halo concentrations. 

Our analysis has revealed that also the halo orbital structure is a subject to Cosmic Web nurturing. 
Especially, for halos with $\vmax\leq80\kms$ (or $M_{200}\simlt 6\times10^{10}\Msun$) the effect of 
a systematic shift in $\rmax$ values at fixed $\vmax$ is clearly visible. For filament and voids populations this 
is a significant result, that can be regarded as an additional environmentally-induced bias.

The signal we found for the halo shape and spin indicate that this quantity for massive halos must be
mostly shaped by the local tidal field. This is indicated by the fact, that only the most massive halos 
in wall and void samples showed a trend with the Cosmic Web. These halos are large enough to experience 
edge-to-edge changes in the local tidal fields, which as a result
can torque and compress the halo. The low-mass halos are too small compared to the typical external tidal field variation scale,
and as such can be seen nearly as point-particles.

In contrast, the concentration and formation redshift seems to be rather unaffected by the local tidal fields. 
This is reflected by the fact that halos above the $M_{th}$ mass threshold have 
the $c(M)$ and $z_{1/2}(M)$ relation (the only exception is node halos).
It remains to be tested,
whether the latter is indeed some kind of a new universal mass-scale at which the environmental effects
become important for the halo and galaxy formation physics. We suspect the value of this threshold,
found by this study to be $M_{th}=6.12\times 10^{10}\Msun$ is not universal. The simulation details,
such as mass and force resolution, together with the assumed cosmological parameters affects both
the precision to which we can resolve the halos and their internal structures as well to the accuracy 
to which the \nexus+ and similar Cosmic Web identification schemes operates. It is very likely, that
$M_{th}$ will be affected by varying cosmology and (hopefully to lesser extent) simulations specifics.
The value we found is however large enough that we can be sure that is not affected by numerical resolution.
Similarly we resolve the Cosmic Web at $0.275\hmpc$ which is 4-times larger than a virial radius
of a $M_{200}=M_{th}$ halo. A separated dedicated study would be required that would offer a closer look
at $M_{th}$ and its variation.

The dependence of halo properties with environment
for low-mass objects is large enough to induce, if ignored, potentially significant systematic biases.
Thus, it needs to be taken into account when interpreting the data and comparing with $\lcdm$ predictions. 
This can be especially important for samples containing low-brightness galaxies that are hosted by such low-mass halos.
On the other hand, one can also use our findings to construct more physically motivated galaxy
formation models  that potentially could lead to
a better agreement between observations and theoretical prediction in the in the small-galaxy regime.

\section*{Acknowledgments}
We are grateful to our colleagues: Aaron Ludlow, Mark Lovel, Sownak Bose and Maciej Bilicki, whose comments on 
the early version of the draft 
were very helpful. We have also benefited from discussions with Carlos S. Frenk at the very early stage of this project.
WAH is supported by the Polish National Science Center
grant no. UMO-2018/30/E/ST9/00698 and UMO-2018/31/G/ST9/03388. 
MC acknowledges the support of the EU Horizon 2020 research and innovation programme 
under a Marie Sk{\l}odowska-Curie grant agreement 794474 (DancingGalaxies)
This project has also benefited from numerical computations performed 
at the Interdisciplinary Centre for Mathematical and Computational Modelling (ICM) 
University of Warsaw under grants no GA67-17 and GA65-30. 

\renewcommand{\bibname}{References}
\bibliographystyle{h-physrev-fix}
\bibliography{coco_cweb_halos}

\end{document}